\documentclass[%
 reprint,
 amsmath,amssymb,
 aps,
 prd,
]{revtex4-2}

\usepackage{graphicx}
\usepackage{dcolumn}
\usepackage{bm}
\usepackage{xcolor}

\begin{document}

\preprint{APS/123-QED}

\title{Investigating the light curve variation of magnetic white dwarfs induced by the axion-photon conversion}

\author{Hao-Chen Tian}
 
\author{Zhao-Yang Wang}

\author{Yun-Feng Liang}
 \email{liangyf@gxu.edu.cn}

\author{Zhao-Wei Du}

 \affiliation{Guangxi Key Laboratory for Relativistic Astrophysics, School of Physical Science and Technology, Guangxi University, Nanning 530004, China}

\date{\today}

\begin{abstract}
Axion-photon oscillation refers to the process of mutual conversion between photons and axions when they propagate in a magnetic field. This process depends on the strength of the background magnetic field, and magnetic white dwarfs provide a natural laboratory for testing this process. In this work, we study the behavior of axion-photon oscillation near magnetic white dwarfs: as the magnetic white dwarf rotates, its magnetic field structure rotates accordingly, causing a periodic change of the magnetic field along the path of photons. These variations affect the axion-photon oscillation process experienced by the photons emitted from the white dwarf, thereby inducing a periodic modulation in the intensity and polarization of the white dwarf's thermal emission that we observe. Our study focuses on the impact of axion effects on the observed light curve variation and conducts a detailed investigation through numerical calculations. Using the light curve data of the white dwarf PG1015+014 obtained from the observations by the Jacobus Kapteyn Telescope, which has a photometric precision of $\sim1\%$, we derive the constraints on axion parameters. In the axion mass range of $\lesssim10^{-8}\,{\rm eV}$, the 95\% credible interval upper limit of the axion-photon coupling $g_{a\gamma\gamma}$ is constrained to $<8.1 \times 10^{-12} \mathrm{GeV^{-1}}$.

\end{abstract}


 \maketitle


\section{introduction}
The axion stems from the strong CP problem in QCD. The Peccei-Quinn (PQ) mechanism is one of the solutions to the strong CP problem, which requires the introduction of a new symmetry, the $U(1)_{\rm{PQ}}$ symmetry. The spontaneous breaking of this symmetry gives rise to the pseudoscalar boson known as the QCD axion \cite{Peccei_1977(1),Peccei_1977(2),Weinberg_1978,Wilczek_1978,Di_Luzio_2020,Marsh_2016,Di_Luzio_2020,adams2023axiondarkmatter}. Similarly, in theories beyond the Standard Model of particle physics, such as string theory and supergravity, additional $U(1)$ symmetries are introduced, which also lead to corresponding pseudoscalar bosons \cite{Peter_Svrcek_2006,Arvanitaki_2010,gendler2024glimmersaxiverse}. Since these particles share properties similar to those of the QCD axion, they are referred to as axion-like particles (ALPs). Through appropriate production mechanisms like the misalignment mechanism \cite{ABBOTT1983133,DINE1983137,PRESKILL1983127}, QCD axions or ALPs can be abundantly produced in the early universe. 
Due to their weak interaction with baryonic matter, they constitute one of the leading candidates for dark matter. By studying the impact of axions or ALPs on gravity or their interactions with standard model particles, direct or indirect detection of axions becomes feasible, for example, some constraints from laboratory experiments \cite{Ehret2010,Bahre2013, CASTCollaboration2024, Anastassopoulos2017, Salemi2021, Pandey2024, Gramolin2021, Kahn2016, Braine2020, Braine2021} and from astronomical observations \cite{Ajello_2016,Abe_2024,Dessert_2022,manzari2024supernovaaxionsconvertgammarays,CHENG2021136611,Dessert_2019,Dessert_2022-2,Gill2011,Davies2023,Jin-Wei2021,Xiao-Jun2021,Hai-Jun2021,GUO2024138631,Hai-Jun2022,Hai-Jun2023,chen2024searchingaxionlikeparticlesxray,Lin-Qing2024,PhysRevD.97.063009,Liang:2018mqm,Liang:2020roo, Ning:2024ozs,Wang:2024jim} were done. References such as \cite{adams2023axiondarkmatter,ringwald2012exploringroleaxionswisps,Redondo_2011,Ringwald_2014,Stadnik_2017,carosi2013probingaxionphotoncouplingphenomenological,RevModPhys.93.015004} provide detailed summaries of various experiments and observations for detecting axions. Since the content described in this paper is applicable to both QCD axions and ALPs, for convenience, the term "axion" hereinafter represents both QCD axions and ALPs without discrimination.

One of the significant properties of axion is the coupling with photon, and axion-photon oscillation is one of the interaction modes. Under a background magnetic field perpendicular to the direction of propagation, axions can transform into photons with polarization parallel to the magnetic field. Likewise, photons with polarization parallel to the magnetic field can convert into axions. During propagation, these two conversions occur alternately, forming an oscillation \cite{Raffelt_1988}. By observing the influence of axion-photon oscillation on photons, the existence of axions could be detected. The strength of the coupling is proportional to the background magnetic field strength, therefore, strong magnetic field environments such as those around magnetic white dwarfs and neutron stars become ideal settings for examining axion-photon oscillation and detecting axions. For example, the thermal radiation from the surface of a compact star propagates within its magnetic field and undergoes axion-photon oscillation, since the oscillation strength depends on the magnetic field strength, photons convert into axions in the strong magnetic field region near the compact star's surface, as they propagate to regions farther from the compact star where the magnetic field weakens, the coupling between photons and axions diminishes, consequently, the axions resulting from the photon conversion cannot completely convert back, leading to alterations in the final electromagnetic radiation that we observe. Axions from the compact star will also convert into photon like this way. This enables us to detect the presence of axions. Relevant studies include Refs. \cite{Gill2011,Dessert_2022,Dessert_2019,Dessert_2022-2,PhysRevD.34.843,PhysRevD.97.123001,Jin-Wei2021,Ning:2024ozs}. In our work, we choose white dwarfs to do the research.

Since only the magnetic field perpendicular to the propagation direction induces axion-photon oscillation, and only photons with polarization parallel to the magnetic field participate in the oscillation, changes in the magnetic field structure will impact the axion-photon oscillation process. Near white dwarfs, as the magnetic white dwarf rotates, its magnetic field also rotates. The specific process of axion-photon oscillation undergoes periodic changes with this rotation, exerting periodic effects on the brightness and polarization of the white dwarf. By verifying whether the light intensity or polarization data within one rotation period conforms to the effects caused by axions, we can indirectly detect axions. 
For a specific telescope, the polarization measurements need to introduce polaroid or zone plates to measure the intensity in different phase differences, which will cause optical loss and introduce extra uncertainties compared to the straightway intensity measurements. Consequently, for the same telescope, if both consider periodic variation, the precision of polarization measurements is often lower than that of photometric measurements. Therefore, in this work, we mainly select light curve variation as the detection method. Note however that the advantage of polarimetry is that its integrated measurement (does not need to be resolved into measurements at different phases) can also reveal information about the existence of axion-like particles (e.g. \cite{Dessert_2022}).

In Sec.~\ref{sec:theory} of this paper, the theoretical details of axion-photon oscillation are elaborated. Sec.~\ref{sec:sec3} constitutes the main body of this work, where the axion-photon oscillation process near magnetic white dwarfs and its influence on light curve variation are calculated. Axion parameters are then constrained using the light curve data of the magnetic white dwarf PG1015+014 from the Jacobus Kapteyn Telescope \cite{Brinkworth_2013}. Sec.~\ref{sec:sec4} presents the discussion and summary of this work.

\section{axion-photon oscillation theory}
\label{sec:theory}
The Lagrangian of the axion field and photon field can be expressed as:
\begin{equation}
    \mathcal{L}=\frac{1}{2}\left(\partial_{\mu}a\partial^{\mu}a-m_a^2a^2\right)-\frac{1}{4}F_{\mu\nu}F^{\mu\nu}-\frac{1}{4}g_{a\gamma\gamma}aF_{\mu\nu}\tilde{F}^{\mu\nu}
    \label{Lagrangian}
\end{equation}
Among them, the first term represents the axion field, where $m_a$ is the mass of axions. The second term is the photon field. The third term is the photon-axion coupling term, in which $g_{a\gamma\gamma}$ is the coupling constant used to describe the coupling strength between axion and photon. Assuming a very high particle occupation number (this assumption is reasonable in the context of our study), the quantization of the field can be ignored and the Lagrangian can be directly substituted into the Euler-Lagrange equation. 
Under the Weyl gauge and by making the WKB approximation (i.e. the wavelength is much smaller than the scale over which the magnetic field changes significantly) and assuming that the axion velocity is near the speed of light (given that the aixon mass is very small ($\lesssim 10^{-6}\,{\rm eV}$) and we are focusing on the observations in the optical band ($\sim 1\,{\rm eV}$), the axion velocity is close to the speed of light), the axion-photon oscillation equation can be obtained \cite{Raffelt_1988}:
\begin{equation}
    \left[ i\partial_x +\omega+
    \left(
    \begin{array}{ccc}
        0 & 0 & 0 \\
        0 & 0 & \Delta_B \\
        0 & \Delta_B & \Delta_a
    \end{array}
    \right) \right]
    \left(
    \begin{array}{c}
        A_\perp \\
        A_\parallel \\
        a
    \end{array}
    \right) = 0
    \label{oscillation equation 1}
\end{equation}
Herein: The $x$-axis points to the propagating direction of the photon or axion, $\omega$ is the energy of the photon or axion, $\Delta_B = \frac{1}{2} g_{a\gamma\gamma}B_t$, where $B_t$ is the strength of the magnetic field component perpendicular to the propagation direction of the photon or axion, $\Delta_a = -\frac{m_a^2}{2\omega}$, $A_\perp$ and $A_\parallel$ are the electromagnetic vector potentials with polarization perpendicular and parallel to $\mathbf{B_t}$, respectively, and $a$ is the field strength of axion. This equation describes the mutual conversion between photon and axion during propagation, which is achieved by the off-diagonal terms $\Delta_B$ in the matrix. Since the off-diagonal terms only exist at positions $(2,3)$ and $(3,2)$ of the matrix, it implies that only photons with polarization parallel to the magnetic field participate in the axion-photon oscillation, while photons with polarization perpendicular to the magnetic field do not participate. This stems from the coupling term $-\frac{1}{4}g_{a\gamma\gamma}aF_{\mu\nu}\tilde{F}^{\mu\nu}$ in Eq.~(\ref{Lagrangian}), where$ F_{\mu\nu}\tilde{F}^{\mu\nu} = \mathbf{E \cdot B}$.

It can be deduced from the oscillation equation that the conversion probability between photons and axions is related to the phase difference between them. Therefore, it is necessary to incorporate the effects that have an impact on the phase difference into the oscillation equation. This implies that in a strong magnetic field environment such as that around a magnetic white dwarf, the QED vacuum birefringence is an effect that cannot be disregarded. To accurately describe this effect, an additional correction term $\frac{\alpha^2}{90m_e^4}\left[\left(F_{\mu\nu}F^{\mu\nu}\right)^2+\frac74\left(F_{\mu\nu}\tilde{F}^{\mu\nu}\right)^2\right]$, i.e. the Euler-Heisenberg term, which originates from the one-loop correction of QED \cite{Raffelt_1988,DUNNE_2005}, needs to be added to the Lagrangian of Eq.~(\ref{Lagrangian}). 
In addition, here we ignore the influence of the photon effective mass caused by plasma $m_A=\sqrt{4\pi\alpha/n_e}$. The reason is that when photons propagate in the interstellar medium, the effective mass is sufficiently small compared to other terms so that it can be seen as zero; and when photons propagate in the atmosphere of the white dwarf, the thickness of the atmosphere is too small to cause significant influence. The reason has also been discussed in Ref.~\cite{Dessert_2022}. After taking these into account in the Lagrangian, the oscillation equation becomes \cite{Raffelt_1988}:
\begin{equation}
    \left[ i\partial_x +\omega+
    \left(
    \begin{array}{ccc}
        \Delta_\perp & 0 & 0 \\
        0 & \Delta_\parallel & \Delta_B \\
        0 & \Delta_B & \Delta_a
    \end{array}
    \right) \right]
    \left(
    \begin{array}{c}
        A_\perp \\
        A_\parallel \\
        a
    \end{array}
    \right) = 0
    \label{oscillation equation 2}
\end{equation}
Where $\Delta_\perp=\frac{4}{2}\omega\left(\frac{\alpha}{45\pi}\right)\left(\frac{B_t}{B_{\rm crit}}\right)^2$, $\Delta_\parallel=\frac{7}{2}\omega\left(\frac{\alpha}{45\pi}\right)\left(\frac{B_t}{B_{\rm crit}}\right)^2$, and $B_{\rm crit}=m_e^2/e$. The introduction of $\Delta_\perp$ and $\Delta_\parallel$ leads to a change in the dispersion relation of photons. The refractive indices of photons with polarization perpendicular and parallel to the magnetic field are different and both become functions of the magnetic field. In a strong magnetic field, this will have a non-negligible impact on the axion-photon oscillation.

If the off-diagonal terms of the matrix are much smaller than the diagonal terms, $\Delta_B$ can be regarded as a perturbation term. It is considered that it does not affect the phase difference of the vector potential and the axion field but only influences the variation of the amplitude. Assuming the initial condition is $\left(A_\perp,A_\parallel,a\right)^T = \left(0,A_0,0\right)^T$, the analytical solution of the equation can be obtained \cite{Dessert_2022}:
\begin{equation}
p_{\gamma \rightarrow a}\left(x\right)=\frac{|a(x)|^2}{|A_0|^2}=\frac{g_{a\gamma\gamma}^2}{4}\left|\int_0^x\,dx'B_te^{-i\int_0^{x'}\,dx''\Delta_{tr}}\right|^2
\end{equation}
Where $p_{\gamma \rightarrow a}(x)$ represents the probability of photons converting into axions, and $\Delta_{tr} = \Delta_\parallel - \Delta_a$. This approximation is applicable in the case when $p_{\gamma \rightarrow a} \ll 1$. Similarly, the conversion from axion to photon follows an analogous process. It is observable that the conversion probability is proportional to the square of the coupling constant; this conclusion will have significant applications in Sec.~\ref{sec:sec3}.

\section{oscillation nearby magnetic white dwarf}
\label{sec:sec3}
Since the thermal radiation from the entire hemisphere of the white dwarf facing the Earth can be observed, it is necessary to integrate the radiation over all positions on the hemisphere visible from Earth. In this work, we define the line connecting the white dwarf and Earth as the polar axis of a spherical coordinate system. The hemisphere of the white dwarf facing Earth is divided into 8100 pixels, with the angles $\theta$ and $\varphi$ incrementing by $2^\circ$ each, calculate the axion-photon oscillation equation corresponding to each pixel of the path separately, and then sum them up. The method is as follows:
\begin{equation}
\begin{aligned}
    I_{\rm total}&=\frac{\iint_S p_{\gamma\rightarrow\gamma}I\cos\theta\,ds}{\pi R_s^2}\\
    &\approx\frac{\displaystyle\sum_{\theta,\varphi}p_{\gamma\rightarrow\gamma}I\cos\theta\cdot R_s^2 \sin\theta\,\Delta\theta\,\Delta\varphi}{\pi R_s^2}
\end{aligned}
\end{equation}
where $p_{\gamma\rightarrow\gamma} = (1 - p_{\gamma\rightarrow a})$ is the photon survival probability along the corresponding path, which is obtained by solving the axion-photon oscillation equation along the path, i.e. Eq.~(\ref{oscillation equation 2}), $I$ is the thermal radiation intensity of the surface of white dwarf, $\Delta\theta = \Delta\phi = \frac{\pi}{90}$, and $R_s$ is the radius of the white dwarf. Due to the temperature gradient between the interior and exterior of the white dwarf as well as a certain degree of transparency, the edge along the line of sight appears darker than the center, a phenomenon known as Limb Darkening, which can be phenomenologically described as $I = I_0 + I_1\cos \theta$ \cite{Euchner_2002,Euchner_2006}. For the axion-photon oscillation of each path, since the magnetic field along the path is not a simple distribution, Eq.~(\ref{oscillation equation 2}) cannot be solved analytically in a straightforward manner. Therefore, numerically solving the equations are required. The thermal radiation emitted from the surface of the white dwarf is unpolarized natural light. To correctly model this, we incoherently superimpose the results of axion-photon oscillations for two beams of left- and right-circularly polarized light with equal amplitudes to simulate the axion-photon oscillation of natural light. For the other three quantities of the Stokes parameters, the calculation method is the same as that of $I_{\rm total}$, and will not be elaborated further. Moreover, the normalized relative light intensity can be obtained:
\begin{equation}
    I_{\rm nor}=\frac{I_{\rm total}}{\iint_SI\cos\theta\,ds/\pi R_s^2}
\end{equation}
It is observable that $I_{\rm nor}$ is equivalent to the average photon survival probability $\langle p_{\gamma\rightarrow\gamma}\rangle$ of the thermal radiation over the entire visible hemisphere of the white dwarf, and thus the average converting probability $\langle p_{\gamma\rightarrow a}\rangle=1-\langle p_{\gamma\rightarrow\gamma}\rangle$.

\subsection{magnetic structure of PG1015+014}

The magnetic white dwarf PG1015+014 was discovered in the Palomar Green survey. Its rotation period is derived to be $98.7 \mathrm{min}$ from the phase-resolved spectroscopic and circular polarization observational data \cite{Wickramasinghe_1988}. In subsequent research, this period is determined with higher precision and fitted a polar
field strength of $B_p = 120 \pm 10\mathrm{MG}$ using a dipole magnetic field model \cite{Schmidt_1991}. Based on their Zeeman spectroscopic dataset, along with flux and circular polarization spectra from each rotation phase of PG1015+014 observed by the VLT at the European Southern Observatory, Ref.~\cite{Euchner_2006} fitted the limb darkening parameters of PG1015+014: the relative light intensity $I = 0.53 + 0.47\cos \theta$, as well as the magnetic field structure and the orientation of the rotation axis. The magnetic field structure was mainly fitted using two models: The first model is a superposition of three individually tilted and off-centered dipole components, while the second model is a truncated multipole expansion of spherical harmonics up to degree $l = 4$. The spectra and circular polarization spectra calculated theoretically based on these two different magnetic field models fitted well with the observation data. Furthermore, they also employed a simple dipole magnetic field model for fitting and obtained $B_p = 131 \pm 1\mathrm{MG}$. It was found that it only fitted well with the observational data at the rotation phase $\phi = 0.25$, but does not match the observational data of the other phases. This implies that the magnetic field of this white dwarf has a more complex structure and cannot be simply described by a dipole magnetic field.

As stated above, given that the magnetic white dwarf PG1015+014 has comprehensive observational data covering its entire rotation period, along with detailed studies of its magnetic field structure, this makes it an ideal candidate for studying the impact of axions on the rotational light variations of magnetic white dwarfs for the indirect detection of axions. Furthermore, the surface magnetic field strength, on the order of hundreds of Gauss, is highly favorable for the occurrence of axion-photon oscillations. Hence, this source is extremely well-suited for our investigations.

It should be noted that the magnetic field structure we obtained based on the spherical harmonic multipole expansion coefficients given in Ref.~\cite{Euchner_2006} is inconsistent with the results presented in their paper. The magnetic field strength we obtained is anomalously large by using their coefficients, reaching 7000~MG at the surface of the white dwarf, which is mainly due to the high values of the high-order spherical harmonic expansion coefficients. 
However, the highest magnetic field strength described in their paper is just over 200~MG. This suggests that there might be errors in the spherical harmonic expansion coefficients provided by Ref.~\cite{Euchner_2006}. On the contrary, we can perfectly replicate the magnetic field model of the superposition of three off-centered dipoles. Hence, for this work, we choose to employ only the magnetic field model of the superposition of three off-centered dipoles. The specific magnetic field parameters can be found in Ref.~\cite{Euchner_2006} Table~3.

To calculate the axion-photon oscillation, it is necessary to determine the radius of PG1015+014. Ref.~\cite{Wickramasinghe_1988} suggested that, given the rotation period of the white dwarf PG1015+014 is comparable to the orbital period of certain AM Her systems (interacting binary systems dominated by magnetic white dwarfs), it is highly likely that PG1015+014 is a remnant of such a binary system. Its mass, approximately $0.13M_\odot$, can be estimated from its rotation speed. By considering only the electron degeneracy pressure to counteract gravity, the corresponding radius is derived as $R_s=0.0247R_\odot$.

\subsection{PG1015+014 light curve variation caused by axion}

\begin{figure}[h]
\includegraphics[width=0.48\textwidth]{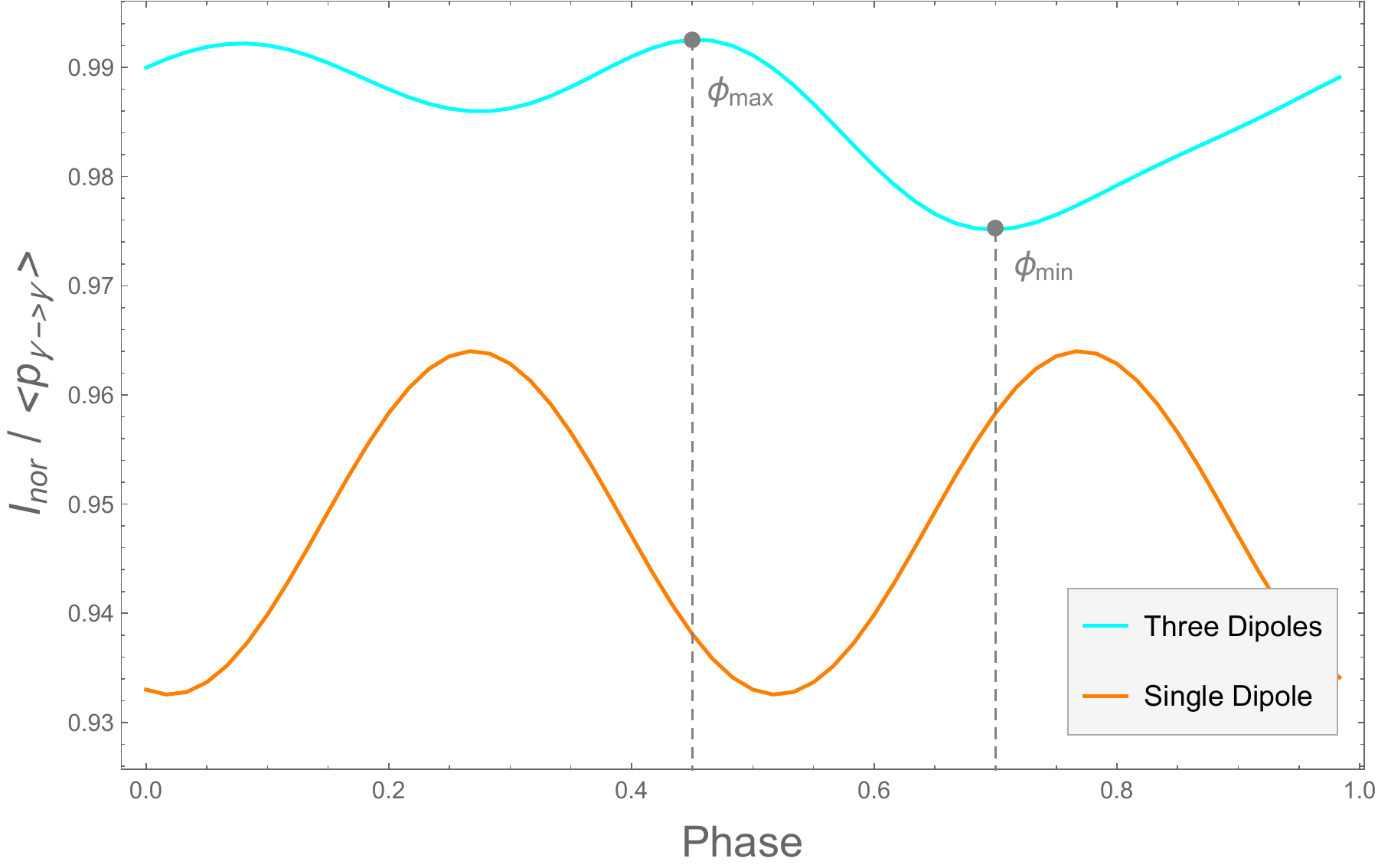}
\caption{\label{light curve} This plot shows the variations of relative emission intensity (equal to the average photon  survival probability $\langle p_{\gamma\rightarrow\gamma}\rangle$) as a function of rotational phase caused by axion effects for the source PG1015+014, for $g_{a\gamma\gamma} = 10^{-11} \mathrm{GeV}^{-1}$ and $m_a = 10^{-8} \mathrm{eV}$. We consider two magnetic field models: (1) the superposition of three off-centered dipoles, and (2) a simple dipole magnetic field. The magnetic field parameters are from Ref.~\cite{Euchner_2006}. $\phi_{max}$ and $\phi_{min}$ represent the phases corresponding to the maximum and minimum light intensities, respectively, for the case of the three off-centered dipole magnetic field model.}
\end{figure}

Given the magnetic field structure and the orientation of the spin axis of PG1015+014, the relative light intensity and polarization due to axion-photon oscillations at each rotational phase can be numerically computed based on the oscillation equation. Taking the axion mass $m_a = 10^{-8}\,\mathrm{eV}$, photon wavelength $\lambda = 790\,\mathrm{nm}$, and coupling constant $g_{a\gamma\gamma} = 10^{-11}\,\mathrm{GeV}^{-1}$ as an example, the variation of the normalized relative light intensity $I_{\rm nor}$ as a function of the rotational phase $\phi$, due to axion-photon oscillation, is shown by the cyan curve in Fig.~\ref{light curve}.

\begin{figure}[h]
\includegraphics[width=0.48\textwidth]{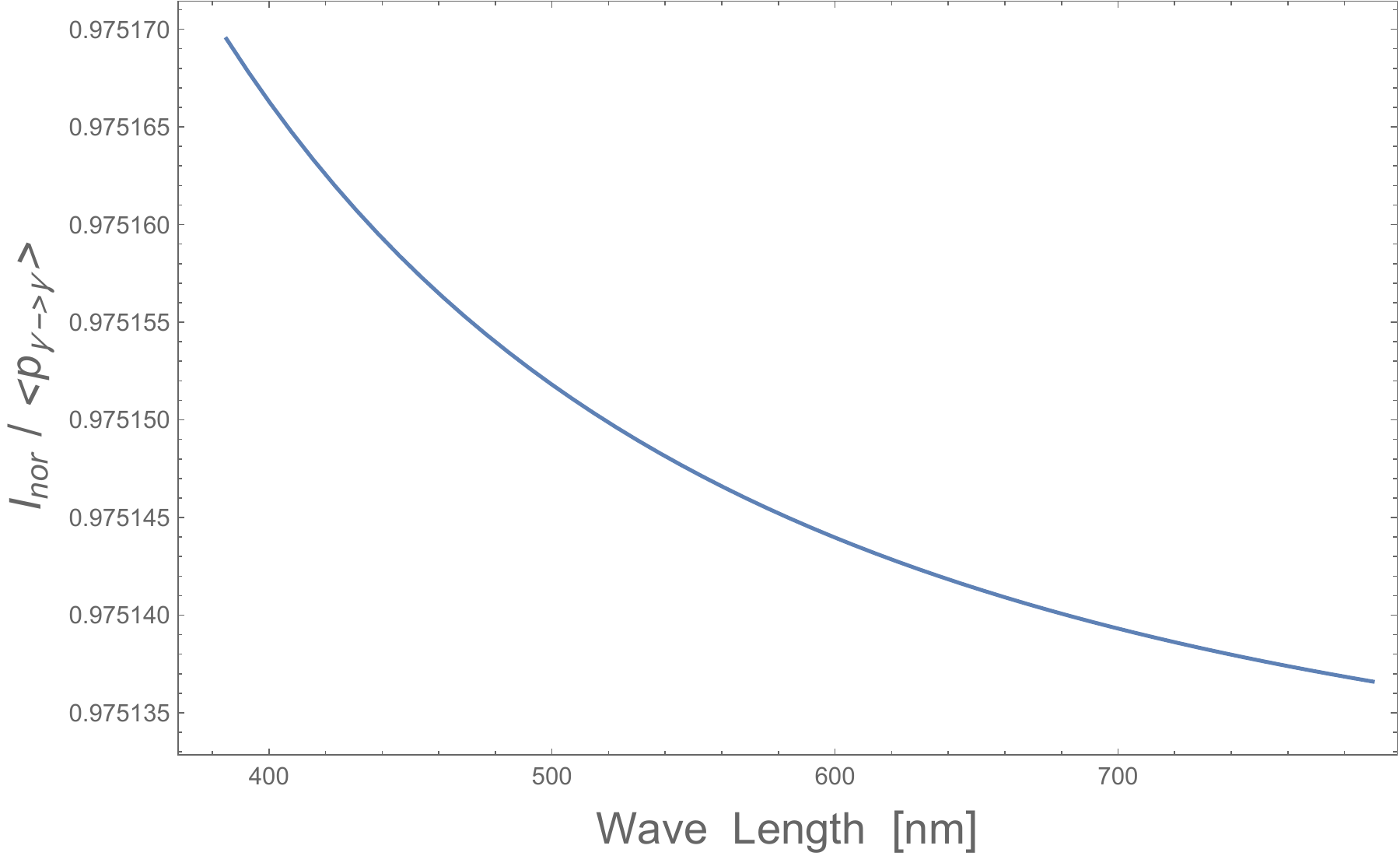}
\caption{\label{wavelength} The normalized light intensity, $I_{\rm nor}$ (equal to the average photon  survival probability $\langle p_{\gamma\rightarrow\gamma}\rangle$) at $\phi_{min} = 0.7$ as a function of wavelength in the three off-centered dipole superposition magnetic field model.}
\end{figure}

Taking the phase with the minimum relative light intensity ($\phi_{min} = 0.7$) as an example, and keeping all parameters except the wavelength unchanged, the relationship between the normalized light intensity and wavelength is shown in Fig.~\ref{wavelength}. It is evident that, although the light intensity decreases as the wavelength increases, changes in wavelength within the visible light band have a negligible effect on the axion-photon oscillation. This indicates that the result for $\lambda = 790\,\mathrm{nm}$ is representative of the axion-photon oscillation behavior for all other wavelengths in the visible light band.

\begin{figure}[h]
\includegraphics[width=0.48\textwidth]{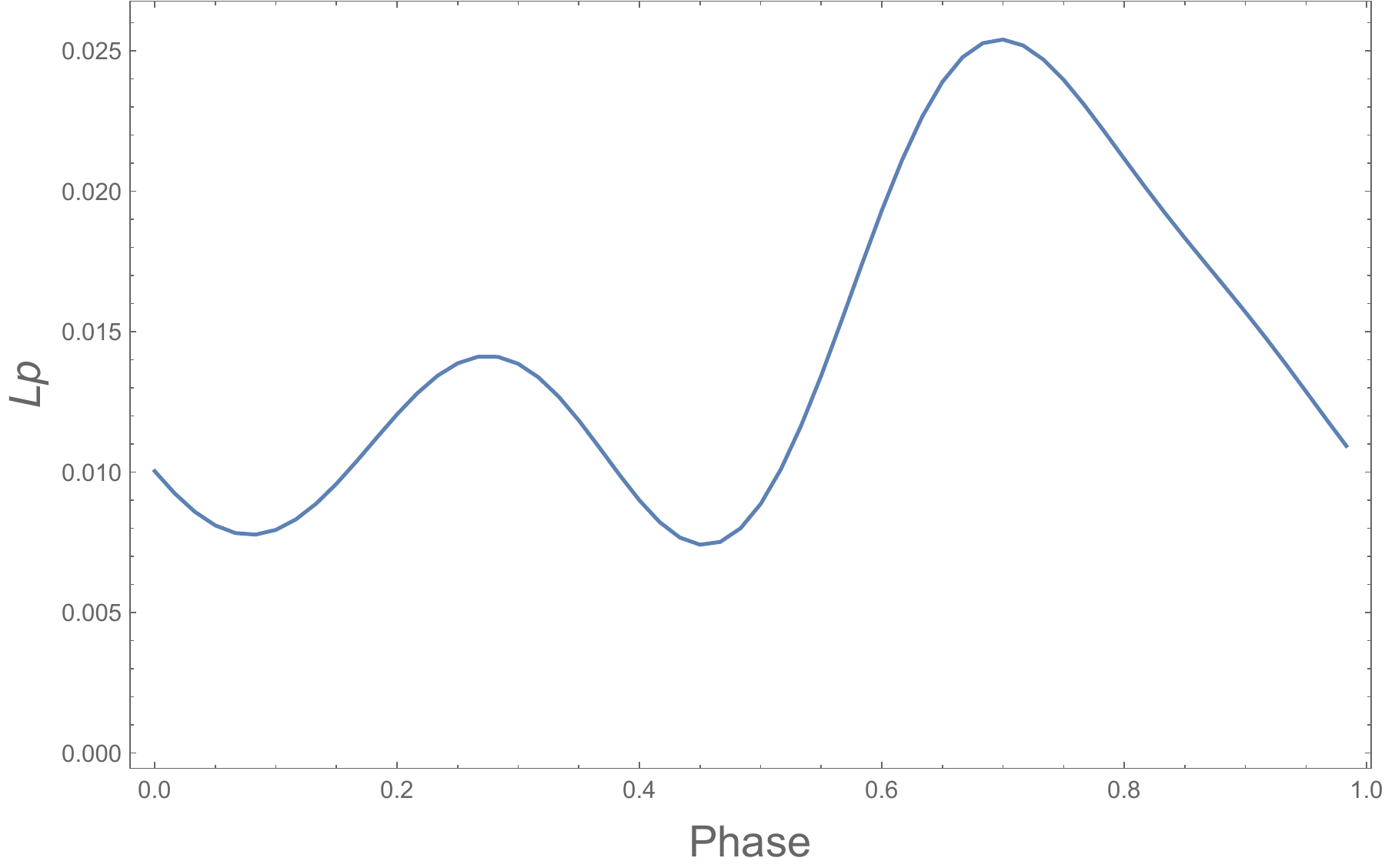}
\caption{\label{Lp}The variation in the degree of linear polarization caused by axion-photon oscillation as a function of the rotational phase in the three off-centered dipole magnetic field model.}
\end{figure}

Meanwhile, we also compute the variation in the degree of linear polarization caused by axion-photon oscillation as a function of the rotational phase, as shown in Fig.~\ref{Lp}. It can be observed that, in such a complex magnetic field scenario, the variation in linear polarization is scarcely different from that of the relative light intensity. 
However, as mentioned above, due to the introduction of polarizers, under the same conditions (same source, same telescope aperture, same exposure, etc.), polarization measurements usually have lower precision than photometric measurements.
The available observational data for light intensity measurements are also typically more plentiful than for polarization. Therefore, if both consider the variation with phase, using light intensity measurements might be more advantageous compared to polarization measurements.
Therefore, in the following, we mainly carry out the analysis based on the variation of light intensity.

\begin{figure}[h]
\includegraphics[width=0.48\textwidth]{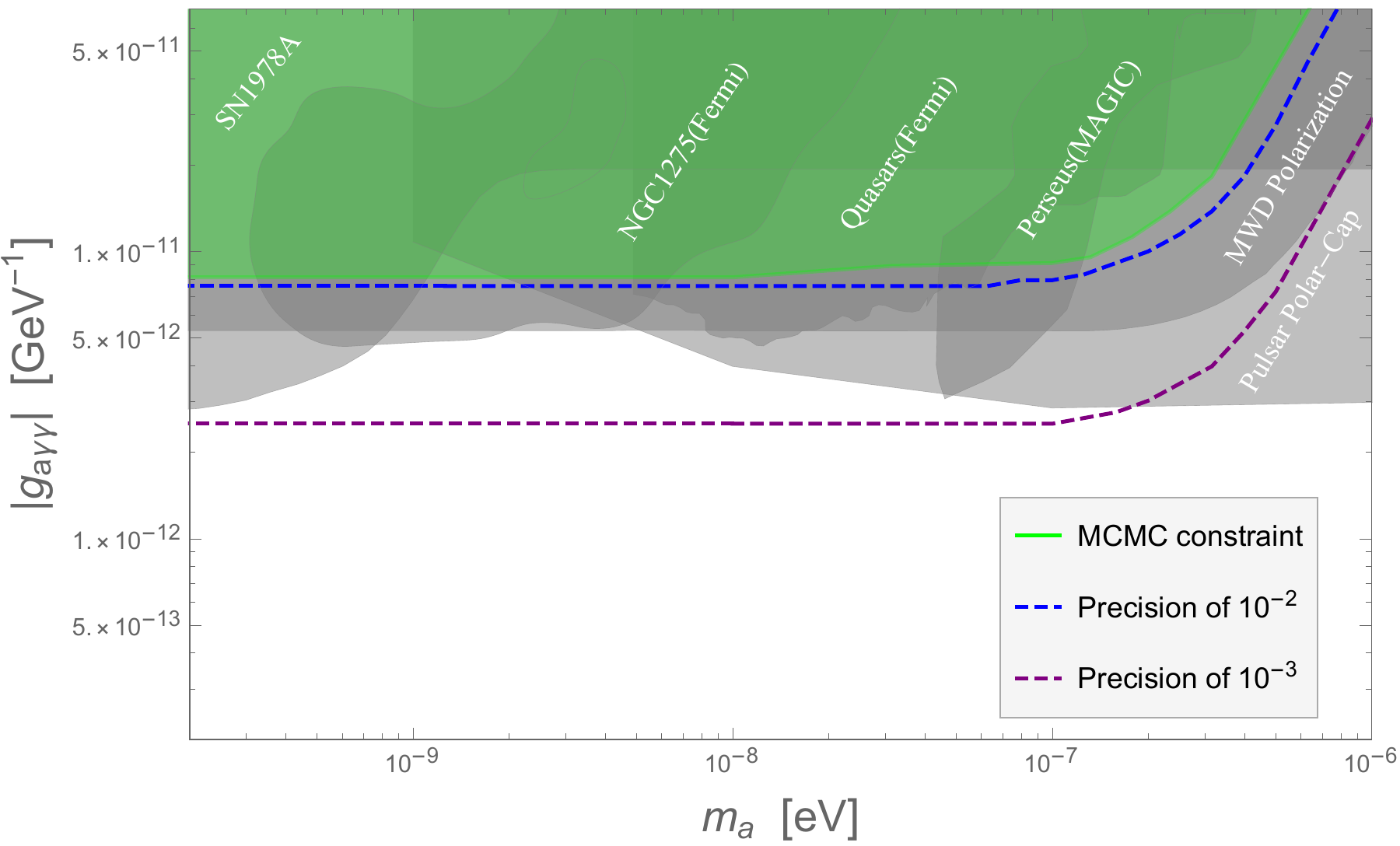}
\caption{\label{exclusion} By adopting the magnetic field structure of a superposition of three dipoles provided by Ref.~\cite{Euchner_2006}, the expected limits on axion parameters with detection precisions of $10^{-2}$ (blue dashed line) or $10^{-3}$ (purple dashed line), together with the exclusion region (green region) obtained through MCMC fitting of the light curve data of PG1015+014 from the Jacobus Kapteyn Telescope \cite{Brinkworth_2013} are shown. The gray areas with thin solid lines represent the exclusion results from other studies \cite{Ajello_2016,Davies_2023,Abe_2024,Dessert_2022,manzari2024supernovaaxionsconvertgammarays}, and their exclusion curve data we used from \cite{Axionlimits}.}
\end{figure}

Assuming that we can resolve a relative light intensity change of $10^{-2}$ or $10^{-3}$ caused by axions in observations, we can estimate the theoretical detection range of this method by comparing the relative light intensity differences between $\phi_{max}$ and $\phi_{min}$. The corresponding detection range in the parameter space, defined by the axion-photon coupling constant $g_{a\gamma\gamma}$ and the axion mass $m_a$, is depicted by the blue and purple dashed lines in Fig.~\ref{exclusion}.

\begin{figure}[h]
\includegraphics[width=0.48\textwidth]{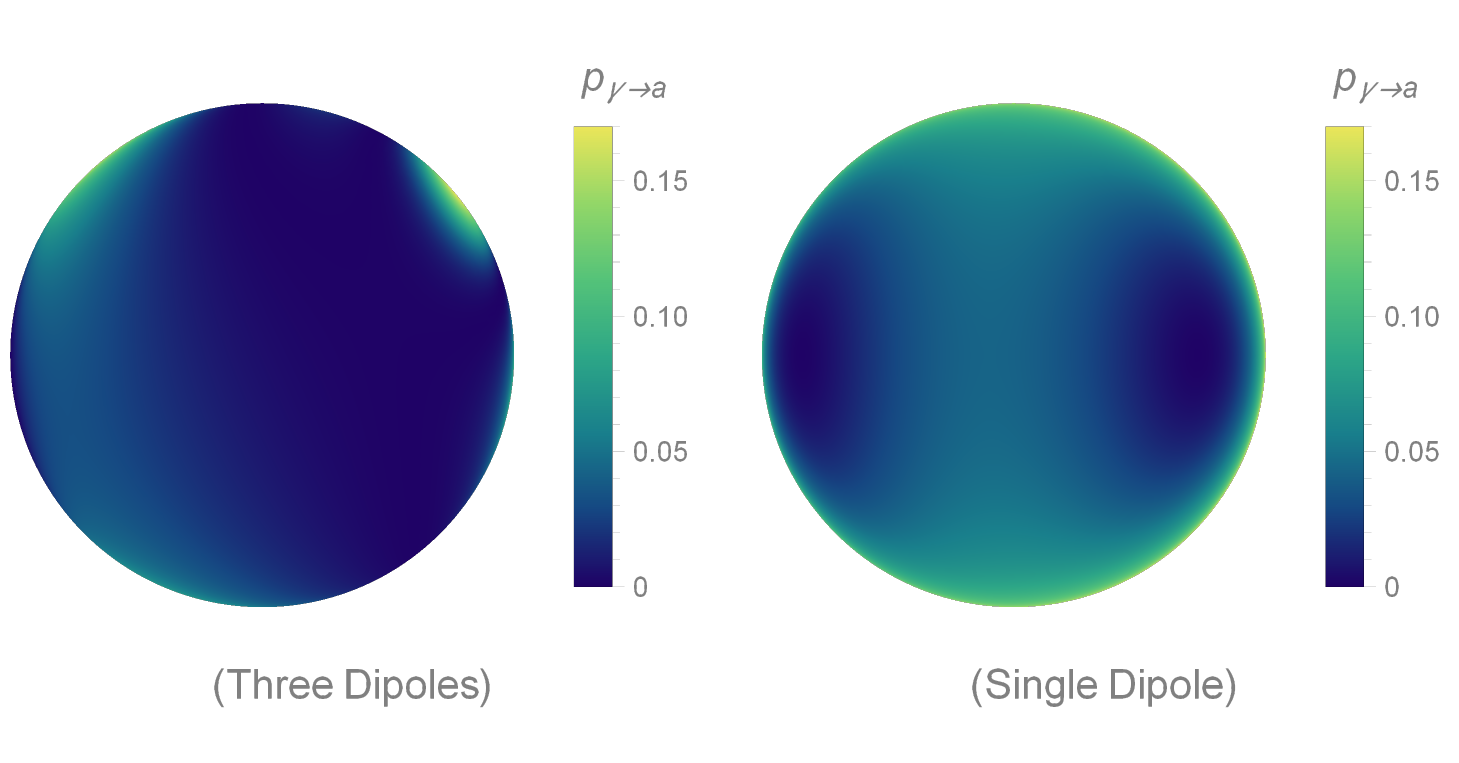}
\caption{\label{DensityPlot} Left panel: The axion conversion probability at various locations on the white dwarf's disk as seen from the line of sight, for the magnetic field structure of a superposition of three dipoles and the rotational phase of $\phi = 0.25$. Right panel: The axion conversion probability at various locations on the white dwarf's disk as seen from the line of sight, for the magnetic field model of a single dipole and the rotational phase of $\phi = 0.25$.}
\end{figure}

Furthermore, Ref.~\cite{Euchner_2006} also presented the fitting results of a simple dipole magnetic field, which only fits the spectrum at the rotational phase $\phi = 0.25$, with the polar field strength $B_p = 131 \pm 1\,\mathrm{MG}$, and the angle between the magnetic axis and the line of sight $\alpha = 83^\circ$. Consequently, we also compute the axion-photon oscillation for this case. Since this magnetic field structure was derived from single-phase observational data and lacks information about the rotation axis, we hypothesize that the rotation axis is perpendicular to the plane formed by the magnetic axis and the line of sight, as this configuration exhibits the most pronounced light variation (the least pronounced variation occurs when the rotation axis coincides with the magnetic axis, because of the symmetry of the dipole magnetic field, it will result in no intensity variation). 
The light curve variation in this scenario is depicted by the orange curve in Fig.~\ref{light curve}, and it can be observed that the light curve variation characteristics in the single dipole magnetic field case differ from the superimposed three dipole model. This distinction is understandable, as the axion-photon oscillation process is highly sensitive to the direction and strength distribution of the magnetic field, leading to different outcomes depending on the magnetic field configuration. To further clarify the differences between the two models, we plot the photon-to-axion conversion probability at various locations across the white dwarf's disk as seen from the line of sight, for both the single dipole and the superimposed three dipole magnetic fields at the rotational phase $\phi = 0.25$, as shown in Fig.~\ref{DensityPlot}.

Since this dipole magnetic field structure fits well only at $\phi = 0.25$ and poorly at the remaining phases, it is deemed unreliable. This raises an important question: do white dwarfs like PG1015+014, whose magnetic field structures significantly deviate from simple dipole configurations, exist widely? If so, this suggests that the results obtained by detecting axions through the magnetic fields of magnetic white dwarfs when these fields are treated simply as dipole fields without adequate consideration of their complexity may not be rigorous.

\subsection{Upper limits on axion-photon coupling $g_{a\gamma}$}

Ref.~\cite{Brinkworth_2013} presented the light curve data of white dwarfs such as PG1015 + 014 obtained by the Jacobus Kapteyn Telescope from August 2002 to May 2003.
The detailed data can be found in Fig.~4 of their paper \cite{Brinkworth_2013}. 
By utilizing this observational light curve, we can constrain the parameters of axions. 

It should be noted that high-magnetic-field white dwarfs experience periodic changes in brightness due to the uneven distribution of magnetic field strength, a phenomenon known as magnetic dichroism. Similarly, for low-magnetic-field white dwarfs, the presence of starspots in their atmospheres can also affect their brightness. Consequently, even in the absence of axion influences, white dwarfs commonly exhibit periodic light variations \cite{Brinkworth_2013}. These light variations, arising from the aforementioned factors, can generally be well-fitted using a trigonometric function (see figures of \cite{Brinkworth_2013} for examples):
\begin{equation}
    I_{\rm bkg}=A+B\,\sin \left( 2\pi f t-\phi^{\prime}\right)
    \label{back curve}
\end{equation}
where $\phi^{\prime}$ represents the initial rotational phase. Since this phase differs from the rotational phase $\phi$ used to describe the magnetic field structure in the actual observational data, they are distinguished here.

Therefore, $\langle p_{\gamma\rightarrow\gamma}\rangle$ itself is not representative of the true brightness variation of white dwarfs. If axions exist, the effect of axions $\langle p_{\gamma\rightarrow\gamma}\rangle$ should be superimposed on the background light variation $I_{\rm bkg}$ caused by magnetic fields and starspots. Thus, the theoretically observable light curve is expressed as:
\begin{equation}
I_{\rm obs}\left(\boldsymbol{\theta}\right)
=I_{\rm bkg}\left(A,B,\phi^{\prime}\right)\times \langle p_{\gamma\rightarrow\gamma}\rangle\left(\phi,g_{a\gamma\gamma}\right)
\label{obs curve}
\end{equation}
where $\boldsymbol{\theta}=\{A,B,\phi^{\prime},\phi,g_{a\gamma\gamma}\}$. 
For PG1015 + 014, the best-fit result with Eq.~(\ref{back curve}) has a reduced-$\chi^2$ of $\sim1.2$, indicating that Eq.~(\ref{back curve}) is a good description of the background null hypothesis at the current observational precision.
Therefore, in our analysis, we assume that the background light variation is described by Eq.~(\ref{back curve}).

To constrain the parameters, the complete light variation for all phases within one cycle needs to be calculated for each set of axion parameters. 
However, fully numerically solving all differential equations are computational expensive (for each case of fixed axion parameters and fixed phases, the oscillation equation needs to be solved $8100\times2 = 16200$ times, where there are 8100 pixels, and the natural light at each pixel is composed of the incoherent superposition of left and right circularly polarized light). Thus, we avoid directly solving the differential equations for axion-induced light variation as above. According to the conclusions of Sec.~\ref{sec:theory}, in the weak mixing approximation, the photon-axion converting probability is proportional to the square of the coupling constant. Render certain that the weak mixing approximation is satisfied, for each mass, we can fix the coupling constant $g_0$ and numerically calculate the average converting probability in each rotational phase $\langle p_{\gamma\rightarrow a}\rangle\left( \phi,g_0 \right)$. Then by using the relationship between the converting probability and the coupling constant, the light variation caused by axions for different coupling constants can be described as:
\begin{equation}
\begin{aligned}
&I_{\rm nor}\left(\phi,g_{a\gamma\gamma}\right)=\langle p_{\gamma\rightarrow\gamma}\rangle\left(\phi,g_{a\gamma\gamma}\right)\\
   &\approx1-\left(\frac{g_{a\gamma\gamma}}{g_0} \right)^2\langle p_{\gamma\rightarrow a}\rangle\left(\phi,g_0\right)
\end{aligned}
\end{equation}

Based on Eq.~(\ref{obs curve}), we obtain the coupling constant constraint corresponding to each axion mass through the Markov Chain Monte Carlo method (MCMC). Assuming the observational errors are Gaussian, the likelihood is:
\begin{equation}
    p\left(\boldsymbol{d}|\boldsymbol{\theta}\right)\propto\prod_{i=1}\frac{1}{\sigma}e^{-\frac{\left(d_i-I_{\rm obs}\left(\boldsymbol{\theta}\right)\right)^2}{2\sigma^2}}
\end{equation}
where $\boldsymbol{d}$ represents the collection of observed light intensity data $d_i$ at each phase, and $\sigma = 0.01$ is the standard deviation of the photometric precision, derived from the error bars in Fig.~4 of Ref.~\cite{Brinkworth_2013}. 
During the fitting/constraint-setting procedure, the parameters of the background sinusoidal function are all free to vary.

\begin{figure}[h]
    \includegraphics[width=0.48\textwidth]{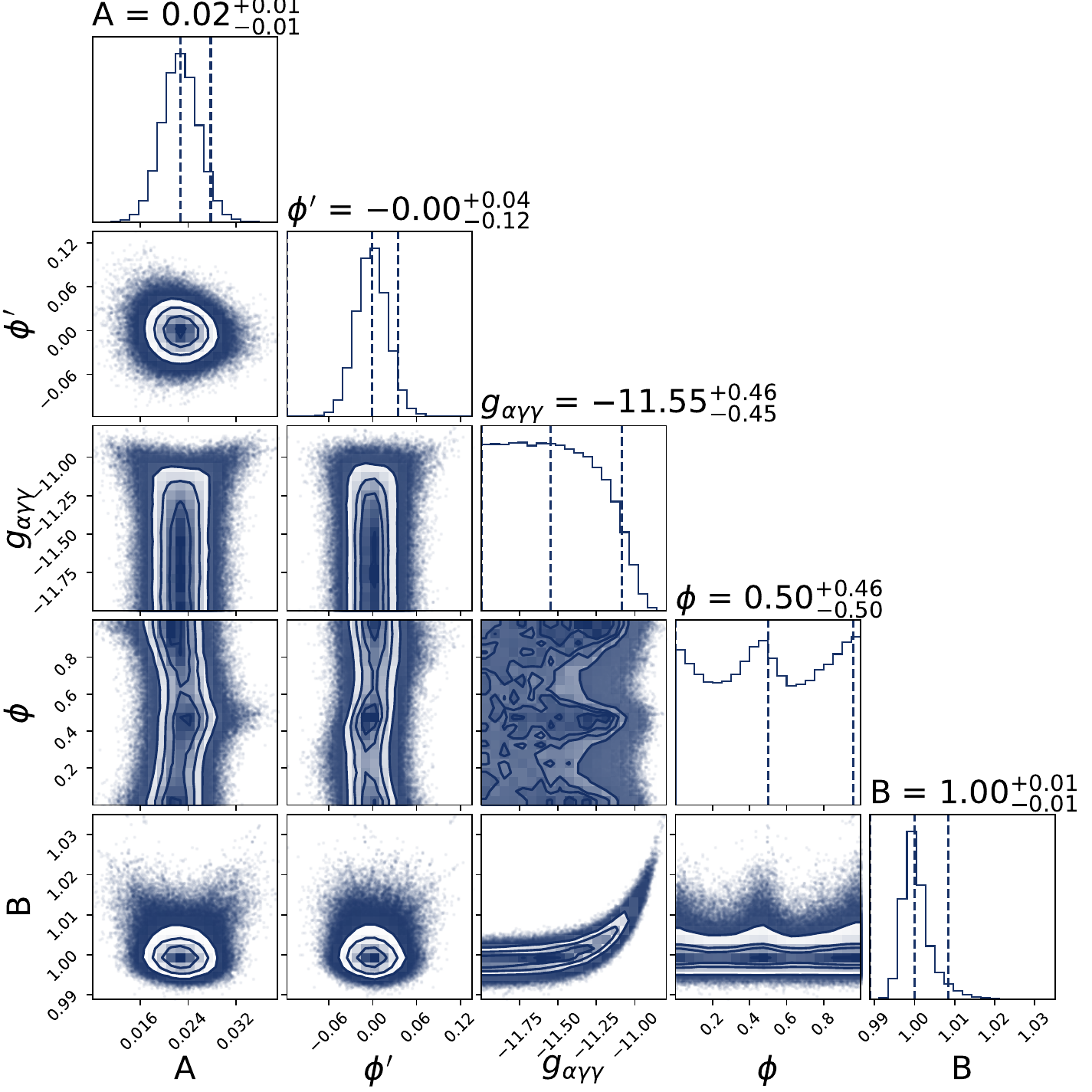}
    \caption{The posterior distribution obtained from MCMC fitting of PG1015+014 light curve data from the Jacobus Kapteyn Telescope \cite{Brinkworth_2013}, assuming an axion mass of  $m_a = 10^{-8} \mathrm{eV}$ and the magnetic field structure of a superposition of three dipoles. The range of $\log_{10}(g_{a\gamma\gamma}) = -11.55^{+0.46}_{-0.45}$ represents the 95\% one-sided credible interval, 
    indicating an upper limit of the coupling constant of $g_{a\gamma\gamma} <8.1\times 10^{-12}\,\mathrm{GeV^{-1}}$.} 
    \label{MCMC result} 
\end{figure}

As an example, in Fig.~\ref{MCMC result} we present the MCMC fitting results for the mass of $m_a = 10^{-8}\,\mathrm{eV}$, by setting the prior for the coupling constant $g_{a\gamma\gamma}$ to a uniform distribution ranging from $10^{-12}\,\mathrm{GeV^{-1}}$ to $10^{-10.5}\,\mathrm{GeV^{-1}}$.
It can be observed that the upper limit of $g_{a\gamma\gamma}$ is constrained to $8.1 \times 10^{-12}\,\mathrm{GeV^{-1}}$ within a 95\% one-sided credible interval. Since the $\Delta_a$ corresponding to $m_a = 10^{-8}\,\mathrm{eV}$ is approximately zero relative to $\Delta_{tr}$, this upper limit applies to the mass range $m_a \in (-\infty, 10^{-8}\,\mathrm{eV}]$ (for example, the result at $m_a = 10^{-9}\,\mathrm{eV}$ shows no difference compared to the result at $m_a = 10^{-8}\,\mathrm{eV}$). The complete exclusion curve covering all values of $m_a$ is shown by the green line in Fig.~\ref{exclusion}. This constraint result is comparable to the expected detection region given earlier with a precision of $10^{-2}$, consistent with the photometric precision standard deviation of the observational data.

\section{discussion and conclusion}
\label{sec:sec4}
The thermal radiation emitted by a magnetic white dwarf undergoes axion-photon oscillation in the presence of strong magnetic fields and may partially convert into axions. As the white dwarf rotates, its magnetic field rotates correspondingly. The magnetic field through which the thermal radiation toward the Earth passes varies periodically, changing the axion-photon conversion probability at different rotational phases. Consequently, this leads to periodic changes in the observed light intensity and polarization. 
This study primarily focuses on the changes in light intensity caused by axions. By calculating the photon survival probability of the white dwarf's thermal radiation at different rotational phases, we can determine the influence of axions on the light curve of white dwarfs. Combining this with the light curve observation data enables us to probe the existence of axions.

The magnetic white dwarf PG1015+014 not only has complete observation data \cite{Brinkworth_2013} for each rotational phase but also has a well-established study of its magnetic structure \cite{Euchner_2006}, making it highly suitable as the specific research object for this work. In this analysis, we primarily use the superposition of three individually tilted and off-centered dipole magnetic fields \cite{Euchner_2006} as our model. We numerically calculate the light curve variation in PG1015+014's thermal radiation caused by axions and present the expected detection regions assuming photometric precisions of $10^{-2}$ and $10^{-3}$. 
If the observation data could reach a photometric precision of $10^{-3}$, this method may provide a stronger constraint on the axion-photon coupling at the masses $m_a \lesssim 10^{-7} \mathrm{eV}$ than the current strongest limit. Additionally, we compare two magnetic field models for PG1015+014: the superposition of three off-centered dipole magnetic fields and a simple dipole magnetic field. It was found that the two models exhibit significantly different characteristics. Since the single dipole magnetic field does not accurately reflect the true conditions of this white dwarf, suggesting that in the works of this type (i.e., those relying on the magnetic fields of magnetic white dwarfs to detect axions), the direct adoption of a dipole magnetic field model may lead to biased results.

By incorporating light curve data from the Jacobus Kapteyn Telescope with a photometric precision of about one percent for PG1015+014, we establish a 95\% credible upper limit of $8.1\times10^{-12}\,\mathrm{GeV^{-1}}$ on the axion-photon coupling constant $g_{a\gamma\gamma}$ in the low-mass range of $m_a\lesssim10^{-8} \mathrm{eV}$ using MCMC fitting. It is noteworthy that due to additional factors such as starspots or uneven magnetic field distribution across the white dwarf's surface, even without considering the presence of axions, the light intensity of white dwarfs varies periodically with rotation. This background light variation can be phenomenologically described using a trigonometric function. This is a relatively coarse way to treat the background, but we believe it is appropriate and effective given the current observational precision. However, if the observational precision is further improved, a simple trigonometric function might not be sufficient to accurately describe the background light variation. Therefore, if one aims to enhance the photometric precision of the observational data to impose stronger constraints on the axion coupling, it becomes essential to develop a more physically grounded and accurate model for the background light variation induced by starspots or uneven magnetic field distribution. 
Although we believe the trigonometric function is a good description of the background light intensity variation at the current observational precision. We should point out that there’s one scenario where this background assumption in our analysis may mask a true signal or overestimate the constraints, i.e., if a non-sinusoidal background combined with an axion effect coincidentally mimics a sinusoidal light curve. But we believe such a coincidence would be rare. Moreover, this issue could likely be avoided by analyzing multiple sources.

In addition, as shown in the joint distribution of $\phi$ and $g_{a\gamma\gamma}$ in Fig.~\ref{MCMC result}, the right boundary reveals an ``$\varepsilon$" shape, and the strength of the constraint on $g_{a\gamma\gamma}$ varies with the change of $\phi$, which is attributed to the unknown of the zero-phase-point of the axion modulation curve. If the zero-phase-point can be determined by analyzing the magnetic field structure, then the uncertainty can be eliminated and the constraint on $g_{a\gamma\gamma}$ can be further improved. In conclusion, detecting axions through the light curve variation of magnetic white dwarf appears to be a promising approach, though there are still some deficiencies and aspects that require improvement. These shortcomings may be addressed in future work.

\begin{acknowledgments}

This work is supported by the National Key Research and Development Program of China (Grant No. 2022YFF0503304) and the Guangxi Talent Program (“Highland of Innovation Talents”).

\end{acknowledgments}

\bibliography{Reference}

\end{document}